\documentclass[
    ,final            % use final for the camera ready runs
  ]
  {aipproc}

\usepackage{amsmath,amssymb}

\layoutstyle{8x11single}

\newcommand{\lapprox}{\ensuremath{\lesssim}}

\begin{document}

\title{Removal of long-lived $^{222}$Rn daughters by electropolishing thin layers of stainless steel}

%alpha decay; Charged-particle spectroscopy, 29.30.Ep; Beta decay, 23.40.-s
%<Replace this text with PACS numbers; choose from this list:
%                \texttt{http://www.aip..org/pacs/index.html}>}

\author{R.W.~Schnee}{
  address={Department of Physics, Syracuse University, Syracuse, NY 13244}
}

\author{M.A.~Bowles}{
  address={Department of Physics, Syracuse University, Syracuse, NY 13244}
}

\author{R.~Bunker}{
  address={Department of Physics, Syracuse University, Syracuse, NY 13244}
%  ,altaddress={<author1 address>} % additional visiting address
}

\author{K.~McCabe}{
  address={Department of Physics, Syracuse University, Syracuse, NY 13244}
%  ,altaddress={<author1 address>} % additional visiting address
}

\author{J.~White}{
  address={Department of Physics, Syracuse University, Syracuse, NY 13244}
%  ,altaddress={<author1 address>} % additional visiting address
}

\author{P.~Cushman}{
  address={School of Physics \& Astronomy, University of Minnesota, Minneapolis, MN 55455, USA}
}

\author{M.~Pepin}{
  address={School of Physics \& Astronomy, University of Minnesota, Minneapolis, MN 55455, USA}
}

\author{V.E.~Guiseppe}{
  address={University of South Dakota, Vermillion, South Dakota 57069, USA}
}

\begin{abstract}
Long-lived alpha and beta emitters in the $^{222}$Rn decay chain on detector surfaces may be the limiting background in many experiments attempting to detect dark matter or neutrinoless double beta decay.
Removal of tens of microns of material via electropolishing has been shown to be effective at removing radon daughters implanted into material surfaces.
Some applications, however, require the removal of uniform and significantly smaller thicknesses. 
Here, we demonstrate that electropolishing $<1$\,$\mu$m from stainless-steel plates reduces the contamination efficiently, by a factor $>100$. 
Examination of electropolished wires with a scanning electron microscope confirms that the thickness removed is reproducible and reasonably uniform.
Together, these tests demonstrate the effectiveness of removal of radon daughters for 
a proposed low-radiation, multi-wire proportional chamber (the BetaCage), 
without compromising the screener's energy resolution.
 %(Zuzel and W?jcik 2012). 
More generally, electropolishing thin layers of stainless steel may 
%be an effective means of removing
effectively remove
 radon daughters without compromising precision-machined parts.
\end{abstract}

\maketitle

%%%%%%%%%%%%%%%%%%%%%%%%%%%%%%%%%%%%%%%%%%%%
%% MAINMATTER
%%%%%%%%%%%%%%%%%%%%%%%%%%%%%%%%%%%%%%%%%%%%

\section{Introduction}
\label{sec:intro}

A particularly dangerous contamination for a number of rare-event searches 
or screening detectors
is the deposition of radon daughters from the atmosphere
onto detector components made of relatively clean materials such as stainless steel.
These radon daughters decay to $^{210}$Pb, a low-energy beta emitter with a long, 22-year half-life, and then to $^{210}$Bi and the alpha-emitting $^{210}$Po.
The $^{210}$Pb daughter is usually plated onto surfaces or, due to the %146\,keV (\,keV) 
recoil energy received from $^{214}$Po (and possibly $^{218}$Po) decay, is implanted into a sub-surface layer of the
material in question ($\lapprox 50$\,nm for stainless steel). 
%It can remain as a main residual contamination after surface treatment
%(cleaning) and may appear after some time through 210Bi (T1/2 = 5.0 d, Emax(b) = 1.2
%MeV) or 210Po (T1/2 = 138.4 d, E(a) = 5.3 MeV).

Electropolishing of stainless steel has been shown to be very effective at removing both $^{210}$Pb and $^{210}$Po~\cite{ZuzelElectropolishSteel2012}.  However, such electropolishing has removed significantly more material ($\sim20$\,$\mu$m) than is allowable for many applications.
In particular, removing such a large thickness would be unacceptable for removing contamination from  
the 25\,$\mu$m-diameter (125\,$\mu$m-diameter) stainless-steel wires constituting the anode (cathode) planes of the BetaCage~\cite{shutt_lrt2004,schneeLRT2006,ahmedLRT2010,protoJINST,LRT2013bunker}. 
 In this paper we present the test results of electropolishing 30--1200\,nm from the surfaces of stainless-steel samples
 that were artificially contaminated by exposing them to a strong radon
 source. 
 %The amount of $^{210}$Po deposited on the stainless steel was measured using alpha spectrometers both before
 %and after cleaning.  
% Past results~\cite{ZuzelElectropolishSteel2012} have indicated that $^{210}$Pb is generally removed more efficiently than  $^{210}$Po.
% As a check, a gamma spectrometer was used to register
%$^{210}$Pb decays for one electropolished sample and one sample that had not yet been electropolished. 
%We also demonstrated the ability to remove a consistent thickness of stainless steel wire using our electropolishing set-up, confirmed by images with a scanning electron microscope. 

\section{Preparation and Electropolishing of Samples}
\label{sect:samples} 

Four unpolished (mill finish) 316 stainless-steel samples, 2 in. $\times$ 2 in. $\times$ 0.1875 in., were cut from steel stock (McMaster-Carr) and prepared.
%50.8\,mm $\times$ 50.8\,mm squares, 3/16 inches thick
The shape and size were chosen in order to fit into the chambers of alpha spectrometers and to allow placement directly on top of the window of a gamma spectrometer. 
Prior to exposure, the samples were scrubbed and rinsed with deionized water followed by a rinse and wipe with alcohol to remove dirt and grease. 
The stainless-steel samples were placed in a chamber 
%how arranged?
and exposed to radon gas between Oct.\ 7, 2011 and Nov.\ 11, 2011 for a total exposure of $5.42\times10^{6}$ Bq\,m$^{-3}$\,day$^{-1}$. 
Measurement in an atomic-force microscope indicated a mean surface roughness of about 8.6\,$\mu$m.
An additional, smoother sample (\#2B finish, $\sim$0.4\,$\mu$m surface roughness), 1.5 in. $\times$ 1.5 in., was exposed under similar conditions in order to explore the dependence of contamination removal on sample surface roughness.
After exposure, the exposed surfaces were not cleaned and all handling was done with gloved hands touching only the sides or bottom of each sample.

\begin{figure}[tb!]
\centering
\includegraphics[height=2.3in]{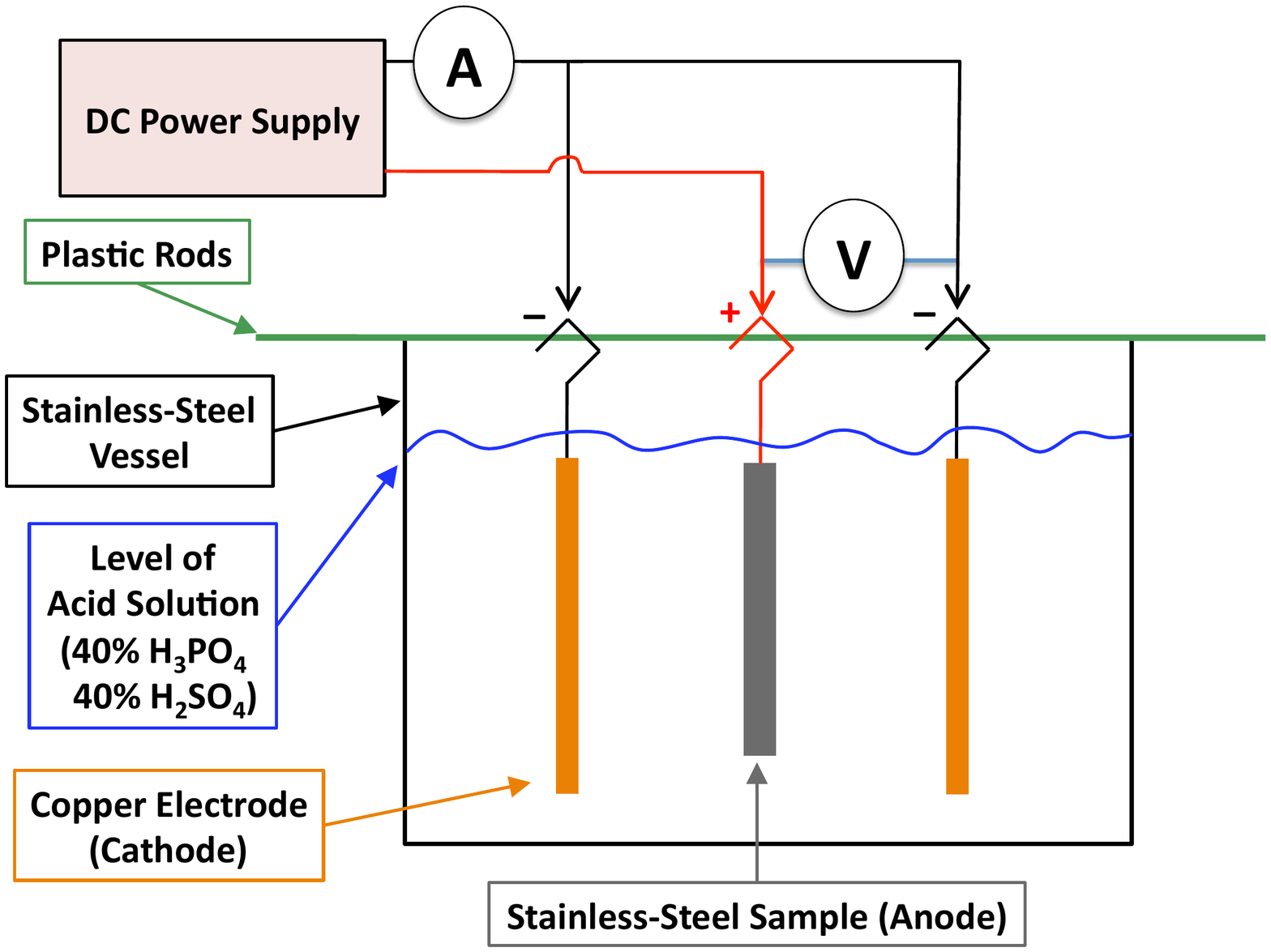}
\includegraphics[height=2.3in]{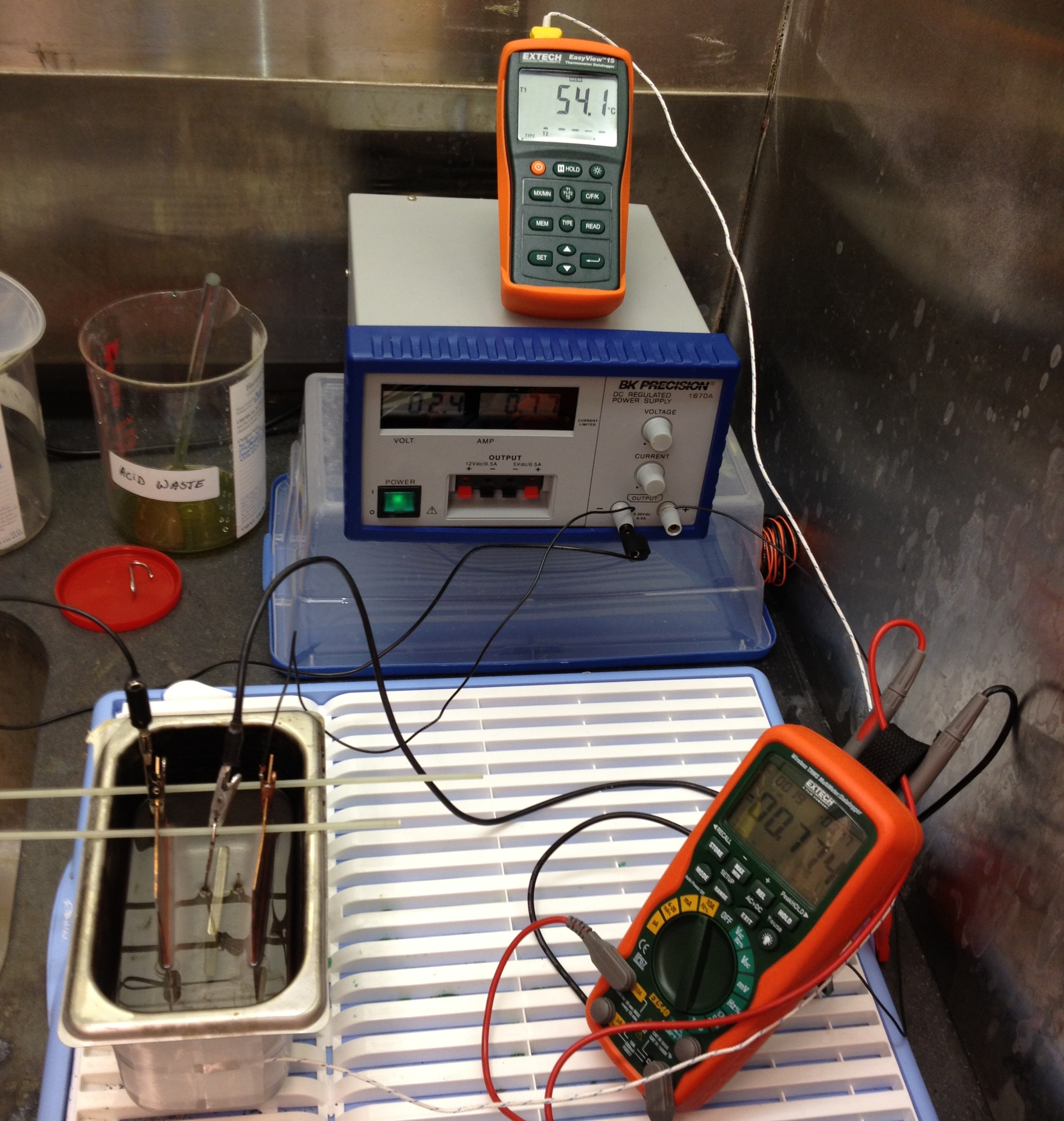}
\caption{{\it Left}: Schematic of the electropolishing setup showing the stainless-steel container and electrode-sample geometry.
{\it Right}: Photo of electropolishing setup within fume hood.}
\label{fig:epsetup}
\end{figure}

The electropolishing was performed using the simple setup shown in Fig.~\ref{fig:epsetup}. It consisted of a
voltage source, current and voltage meters, and two 100\,mm $\times$ 100\,mm square cathodes made out of
copper. The applied electrolyte was a mixture of H$_3$PO$_4$ (40\%) and H$_2$SO$_4$ (40\%). 
The
applied voltage was 2.4 V (DC).
Although the current during electropolishing tends to decrease over time, the value was effectively constant ($\sim1$\,A) during the short (10\,s -- 2\,m) electropolishing runs performed here.
We determined the amount of material removed by weighing the samples with a digital scale (Mettler AE 200) with a precision of $\sim$1\,mg (with its glass doors closed).  Repeated measurements of standards before and after each sample weighing allowed compensation for drifts in the absolute weight scale and estimates of the mass-loss uncertainty.  
Average removal rates, assuming a stainless steel density $\rho = 8.0$\,g\,cm$^{-3}$, were about 4\,nm/s for the square samples. %, and 30\,nm/s for the 25-micron wire.

%\section{Detection Systems}
\section{Measurement of uniformity and removal of contamination}

\begin{figure}[tb!]
\centering
\includegraphics[width=2.0in]{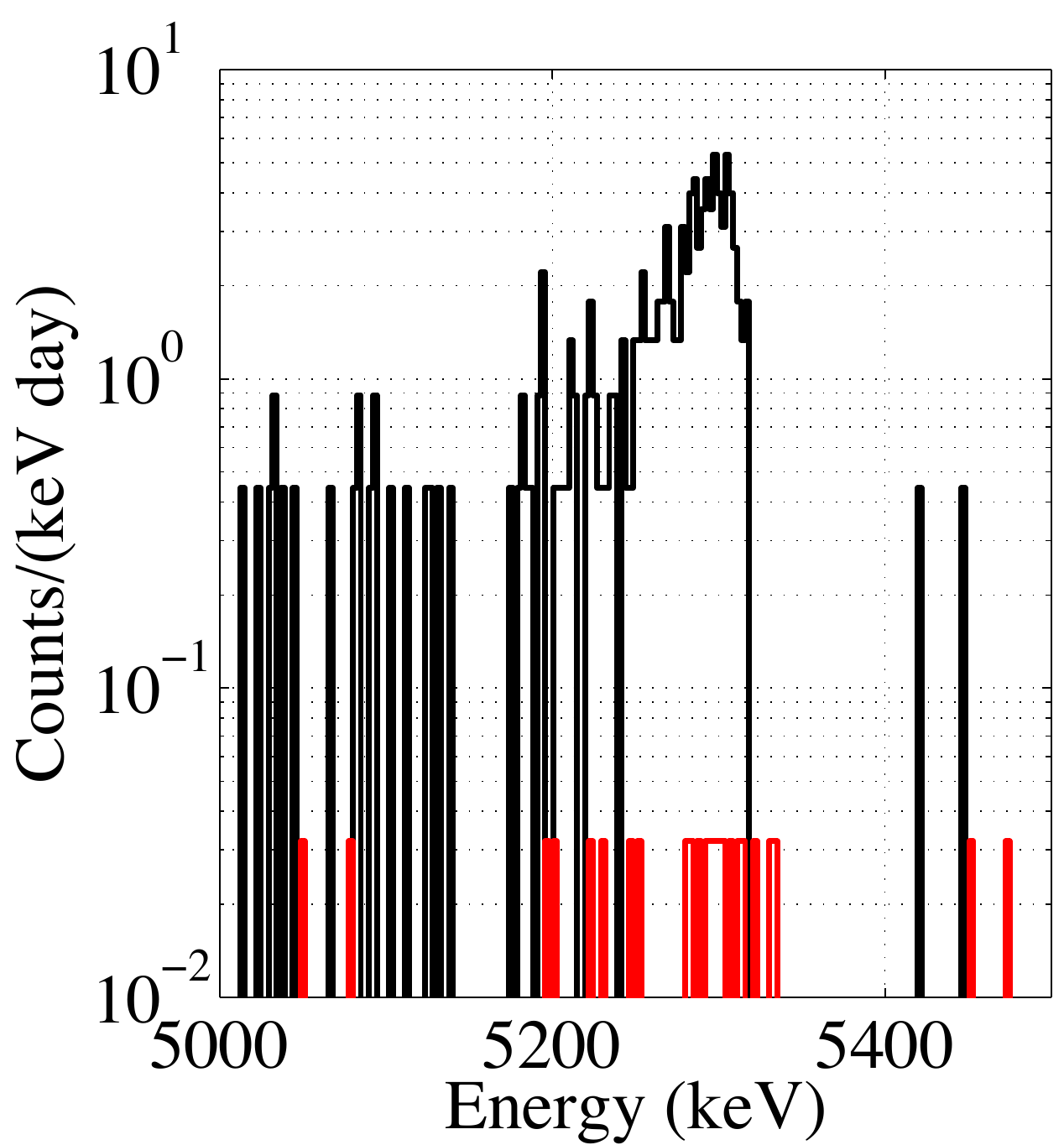}
\includegraphics[width=4.0in]{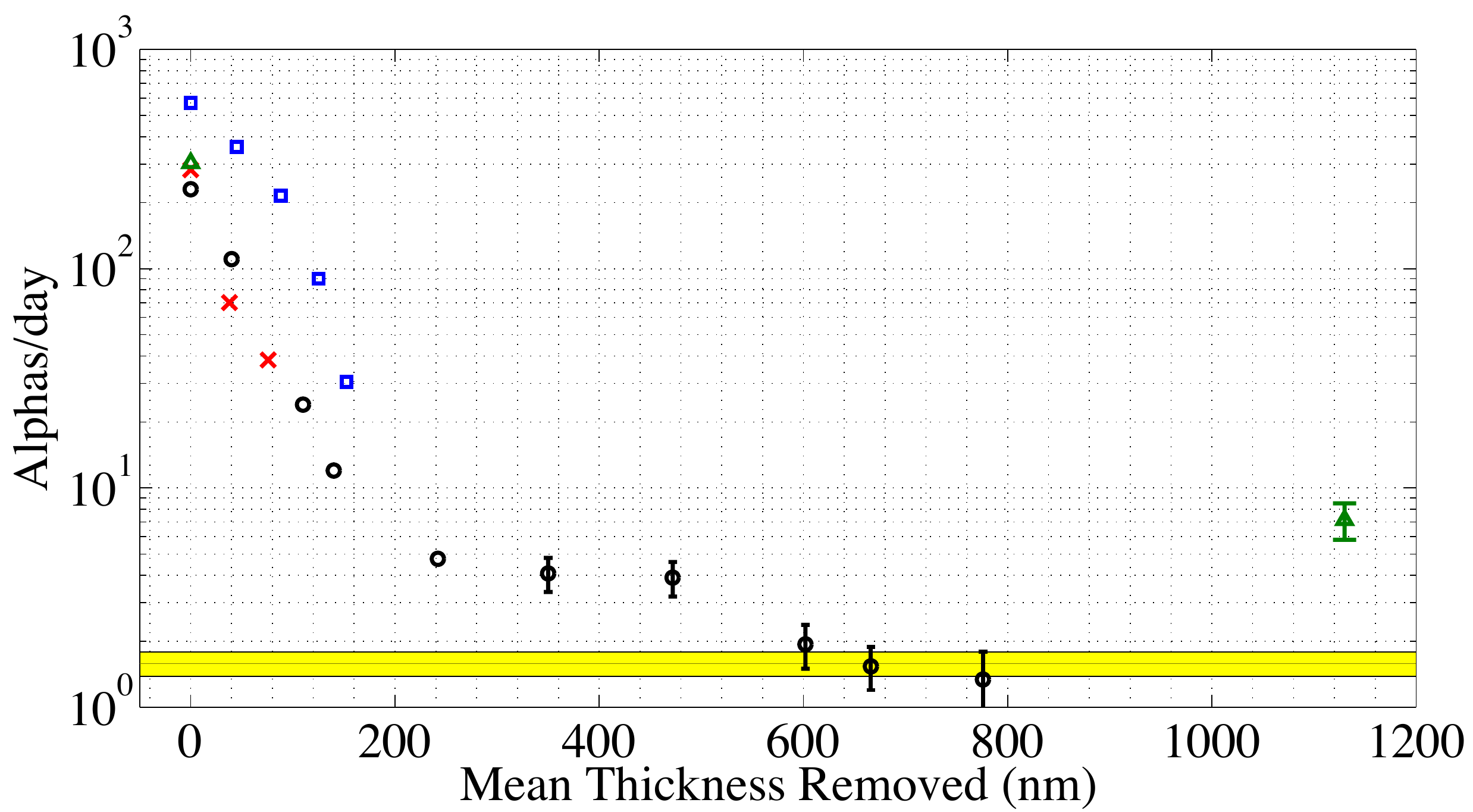}
\caption{{\it Left}: Raw spectra of Sample \#3 both before (black) and after (red) removal of 670\,nm of thickness.  A reduction factor $>100\times$ is apparent for events in the 5.2--5.4\,MeV region of interest for $^{210}$Po.  The rate after removal is consistent with the background rate of the detector (when empty).
{\it Right}:  Alpha detection rates for Sample~\#1 (green triangles), Sample~\#3 (black circles), Sample~\#4 (blue squares), and the sample with the \#2B finish (red $\times$'s), as a function of the mean cumulative thickness of material removed.  Uncertainties are smaller than the symbols if no error bars are shown.  Lack of reduction in alpha rate from 200\,nm to 500\,nm removed from Sample~\#3 is suspected to have occurred due to $^{210}$Po contamination in the rinse water.  After fresh ultra-pure water was used for subsequent electropolishing, the alpha rate dropped to the background rate of the detector (yellow horizontal band).
Preliminary results shown here do not include (small) corrections for drifts in the digital scale, drifts in the alpha identification efficiency due to gain drifts, nor for grow-in of the $^{210}$Po over the time to make these measurements.
}
\label{fig:alphas}
\end{figure}

 An alpha spectrometer (ORTEC Alpha Ultra-AS 33.8\,mm diameter Si detector) with a background of $1.6 \pm 0.2$\,cpd in the energy region of interest from 5.20--5.35\,MeV measured $^{210}$Po surface activity.   A
 small distance between the detector and the sample (5\,mm)  %CHECK!  9mm?
  and low operating pressure
 of 200\,mTorr allow for alpha spectroscopy with a high efficiency (13.8\%) and good energy resolution, so long as the contamination is on the sample surface (see Fig.~\ref{fig:alphas}).  
 The absolute efficiency was taken as the geometrical efficiency of the detector, and 
 checked by comparing the measured rate for a calibrated source to the expected rate based on the calculated geometrical efficiency.
However, since final results depend only on relative changes in the signal
from before to after cleaning, the uncertainty on the absolute calibration of the detector is ignored here.

Figure~\ref{fig:alphas} shows the results of the electropolishing on the measured $^{210}$Po alpha rates, ignoring (small) corrections for drifts in the digital scale, drifts in the alpha identification efficiency due to gain drifts, and for grow-in of the $^{210}$Po over the time to make these measurements.
%Removal of 160\,nm reduced the alpha rates by a factor of about 20.  
The reduction approximately follows an exponential profile with characteristic mean thickness $t \approx 50$\,nm causing a $1/e$ reduction, with similar results for all samples with total rates above 10 alphas/day.  %Below this rate, both Sample~1 (for which no intermediate data were taken) and Sample~3 show a stalling in the reduction of $^{210}$Po .  
For Sample~\#3, reduction stalled with the alpha rate at about 5 events/day until an improved technique, using a fresh solution of ultra pure water, was used.  At that point, the rate on the sample was reduced to a level at or below the chamber background.  Contamination in the rinse water may also affect lowest-rate data taken with Sample~\#1 and the sample with the \#2B finish.
%add sentence saying check to comeÉ last data point on 2b may also be high due to old water

A likely reason why only a fraction of the deposited $^{210}$Po is removed per 50\,nm of material removed is the sample surface roughness. 
%and $^{210}$Po cannot be deposited further than 50\,nm by recoil of the daughter $^{214}$Pb or $^{210}$Pb nuclei from alpha decays.
%One possibility is some diffusion of  $^{210}$Po in steel is occurring, with some atoms working their way deeper into the steel.  
Electropolishing tends to smooth out surfaces by preferentially removing atoms from protrusions, 
so much of the removed material is likely from parts of the sample deeper than any $^{210}$Po.
The first data from the smoother sample suggests contamination was removed at a slightly faster rate.  Studies in the near future will concentrate on determining the relationship between $t$ and surface roughness by electropolishing smoother samples while correcting for systematic effects.

%Although previous results~\cite{ZuzelElectropolishSteel2012} indicate that $^{210}$Pb is removed much more effectively than $^{210}$Po, 

%%%%%%%%%%%%%%%%%%%%%%%%%%%%%%%%%%%%%%%%%%%%
%% Sample figure:
%%
%% The option [height=...] scales the picture to the given height,
%% without it it would be printed at its nominal size
%%%%%%%%%%%%%%%%%%%%%%%%%%%%%%%%%%%%%%%%%%%%

%\begin{figure}
%  \includegraphics[height=.3\textheight]{golfer}
%  \caption{Picture to fixed height}
%\end{figure}
\begin{figure}[tb!]
\centering
\includegraphics[height=1.52in]{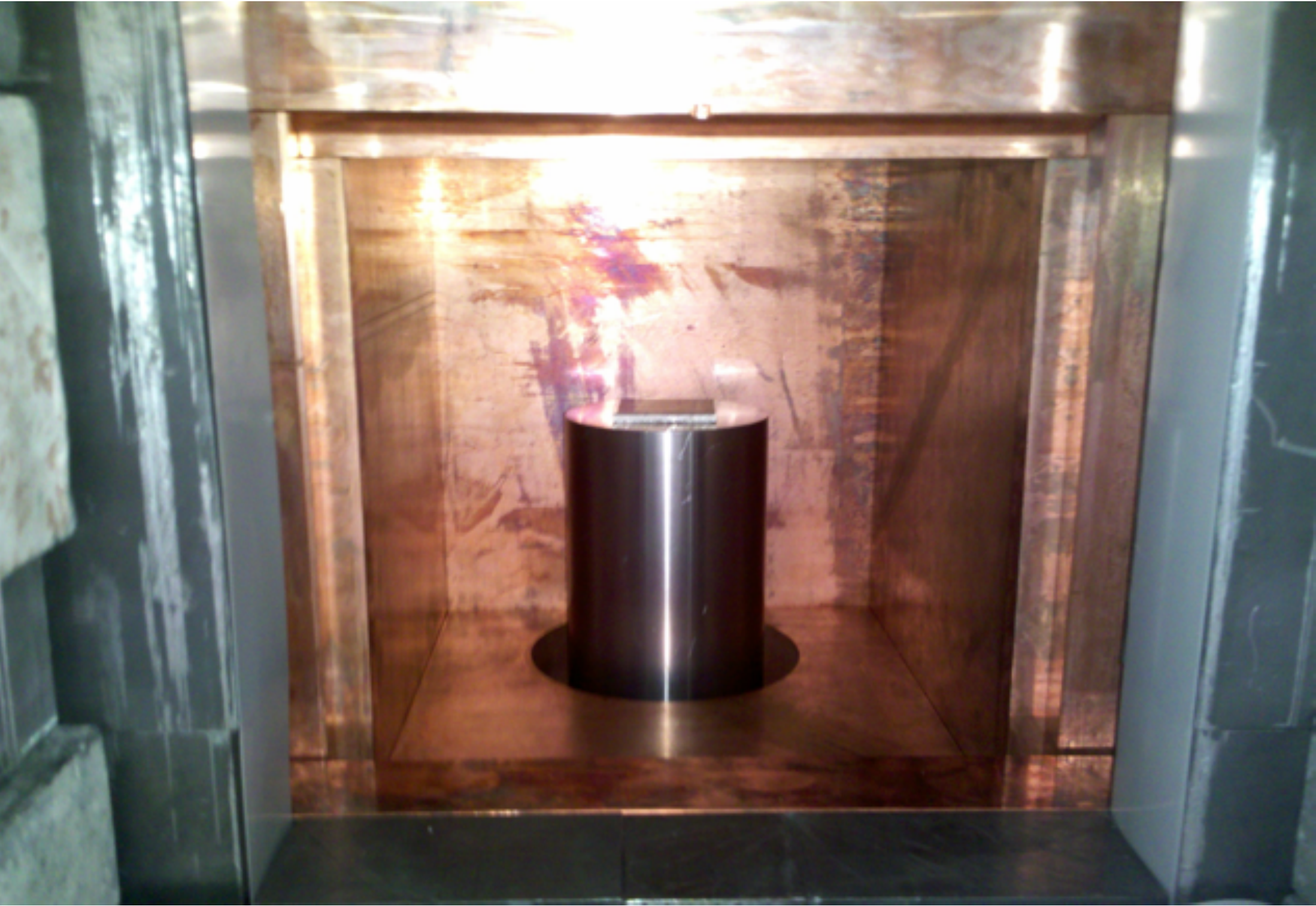}
\includegraphics[height=1.52in]{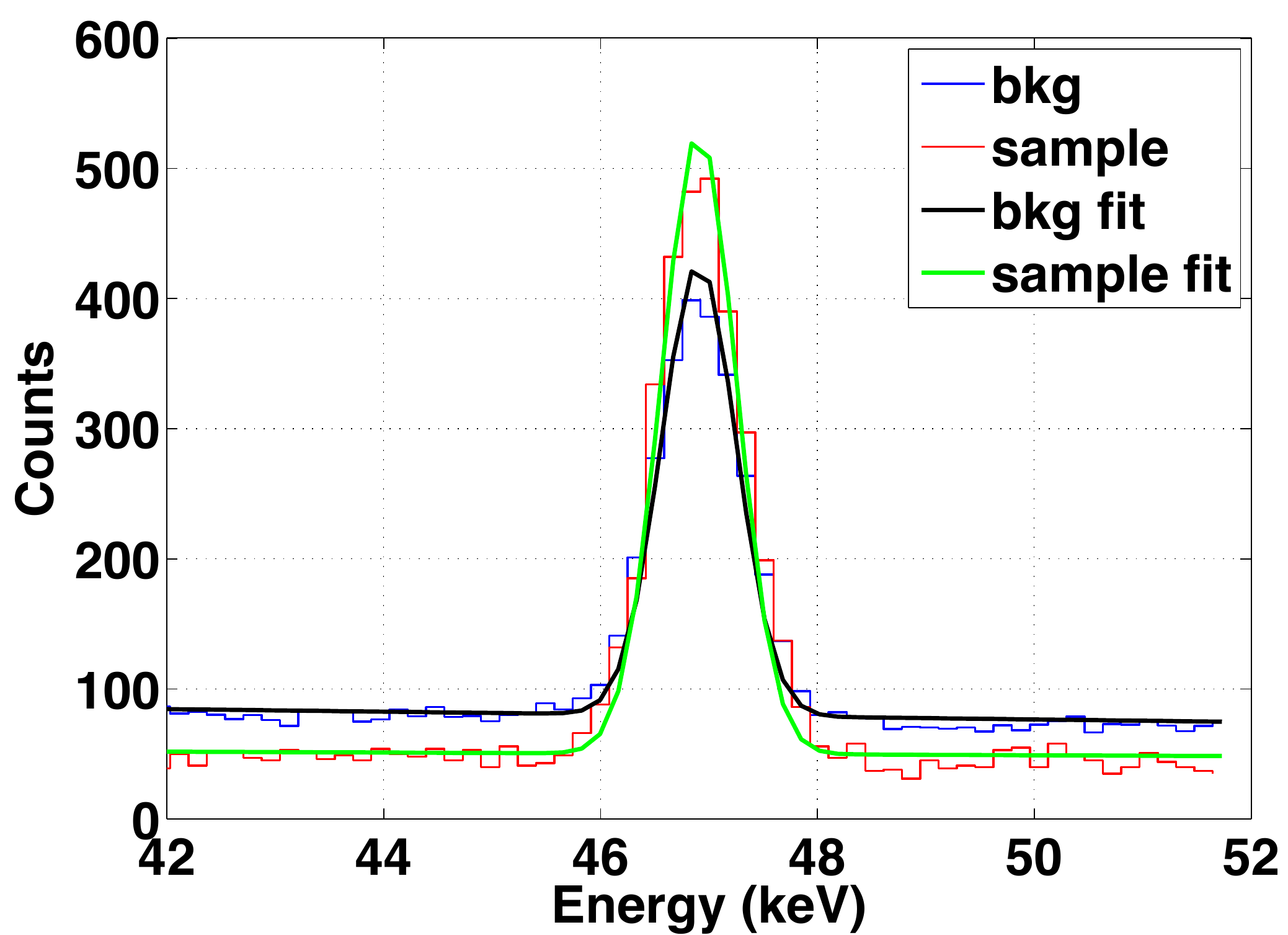}
\includegraphics[height=1.52in]{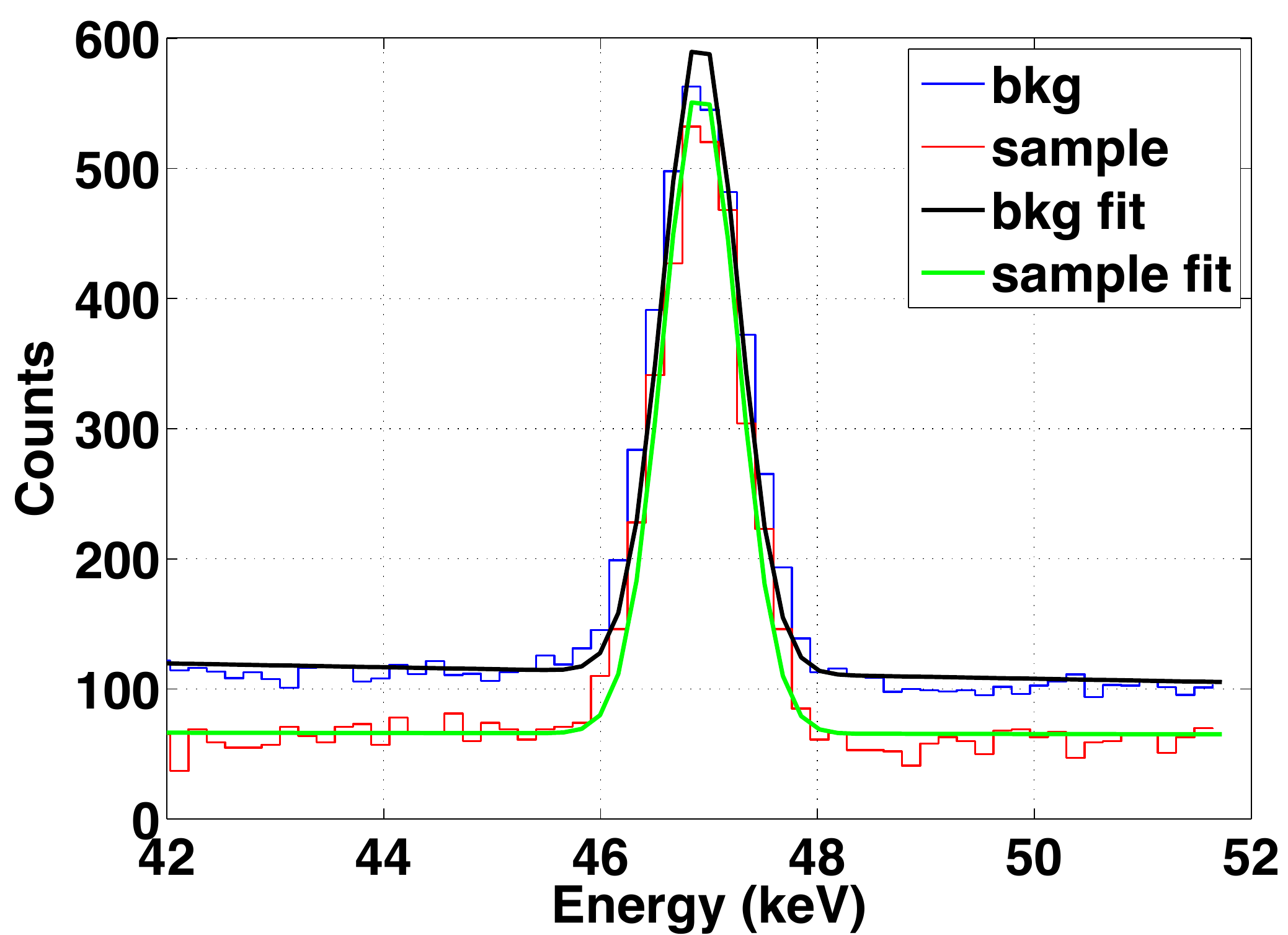}
\caption{{\it Left}: Inside of GOPHER shielding, showing HPGe detector with sample directly on top of the Al window, surrounded by the Cu %shielding 
liner.  {\it Center}: GOPHER spectrum (red histogram) of Sample \#2 (control) with fit (green curve) compared to scaled background sample (blue histogram with black curve showing fit), indicating a $^{210}$Pb activity of $14.6 \pm 1.3$\,Bq/m$^2$.  Combined with alpha counting before electropolishing, this  measurement indicates that the $^{210}$Pb activity for Sample~\#1was $13.5 \pm 1.3$\,Bq/m$^2$ before electropolishing.  {\it Right}: GOPHER spectrum (red histogram) of Sample \#1 with fit (green curve) compared to scaled background sample (blue histogram with black curve showing fit), indicating a $^{210}$Pb activity of $<1.1$\,Bq/m$^2$ after electropolishing 1.2\,$\mu$m off.}
\label{fig:gopher}
\end{figure}

Removal of $^{210}$Pb by electropolishing small thicknesses of steel was confirmed
by observing the 46.6\,keV gamma line with two samples placed directly on top of the 1.6\,mm  aluminum window of the GOPHER n-type high purity germanium detector, as shown in Fig.~\ref{fig:gopher}. 
%For two samples, $^{210}$Pb was identified by observing the 46.6\,keV gamma line with samples placed directly on top of the 1.6\,mm  aluminum window of
% the GOPHER n-type high purity germanium detector, as shown in Fig.~\ref{fig:gopher}.
 GOPHER sits within  a high-purity copper-lined lead shield and includes a radon purge unit with sample load-lock.
 %citation?  more detail on thickness of shield, activity?	
 % (HPGe, 25 \% relative detection efficiency,
 A detailed, custom GEANT4 simulation of the detector sample geometry indicates a 35\% efficiency for 46.6\,keV gammas originating on the surface of the sample facing the detector window.
 % more on background, consistency between backgrounds from both samples
 Samples~\#1 (with 1.1\,$\mu$m removed) and \#2 (control) were counted in the GOPHER HPGe counter for 18 and 14\,days respectively.
Sample~\#2 had $14.6\pm1.3$\,Bq/m$^{2}$ $^{210}$Pb.  This measurement, combined with the relative alpha rates of Samples \#1 and \#2 before Sample \#1 was electropolished, allows the  $^{210}$Pb activity on Sample \#1 before electropolishing to be inferred as $13.5\pm1.3$\,Bq/m$^{2}$.
Measurement of Sample~\#1 after electropolishing was consistent with detector backgrounds, indicating an upper limit at the 90\% confidence level of 1.1\,Bq/m$^{2}$.  The reduction factor due to electropolishing was $\geq 12$, consistent with expectations that the reduction of $^{210}$Pb would be at least as large as  the measured factor of $\sim50\times$ for $^{210}$Po for the same Sample~\#1.

%Measurements with the GOPHER HPGe counter confirmed that electropolishing small thicknesses of steel is also effective at removing $^{210}$Pb.
%Samples~1 (with 1.1\,mum removed) and 2 (control) were counted in the GOPHER HPGe counter for 18 (?) days respectively.
%The main goal in this analysis was to investigate the improvement in the Pb-210 46.6 KeV line between the unpurified and purified samples. A large improvement was seen from the prior samples Pb-210 14620 +/- 1340 mBq/m$^2$ and the upper limit set on the purified sample of 1124 mBq/m$^2$. 

\subsection{Uniformity of electropolished wires}

\begin{figure}[tb!]
\centering
\includegraphics[width=3.2in]{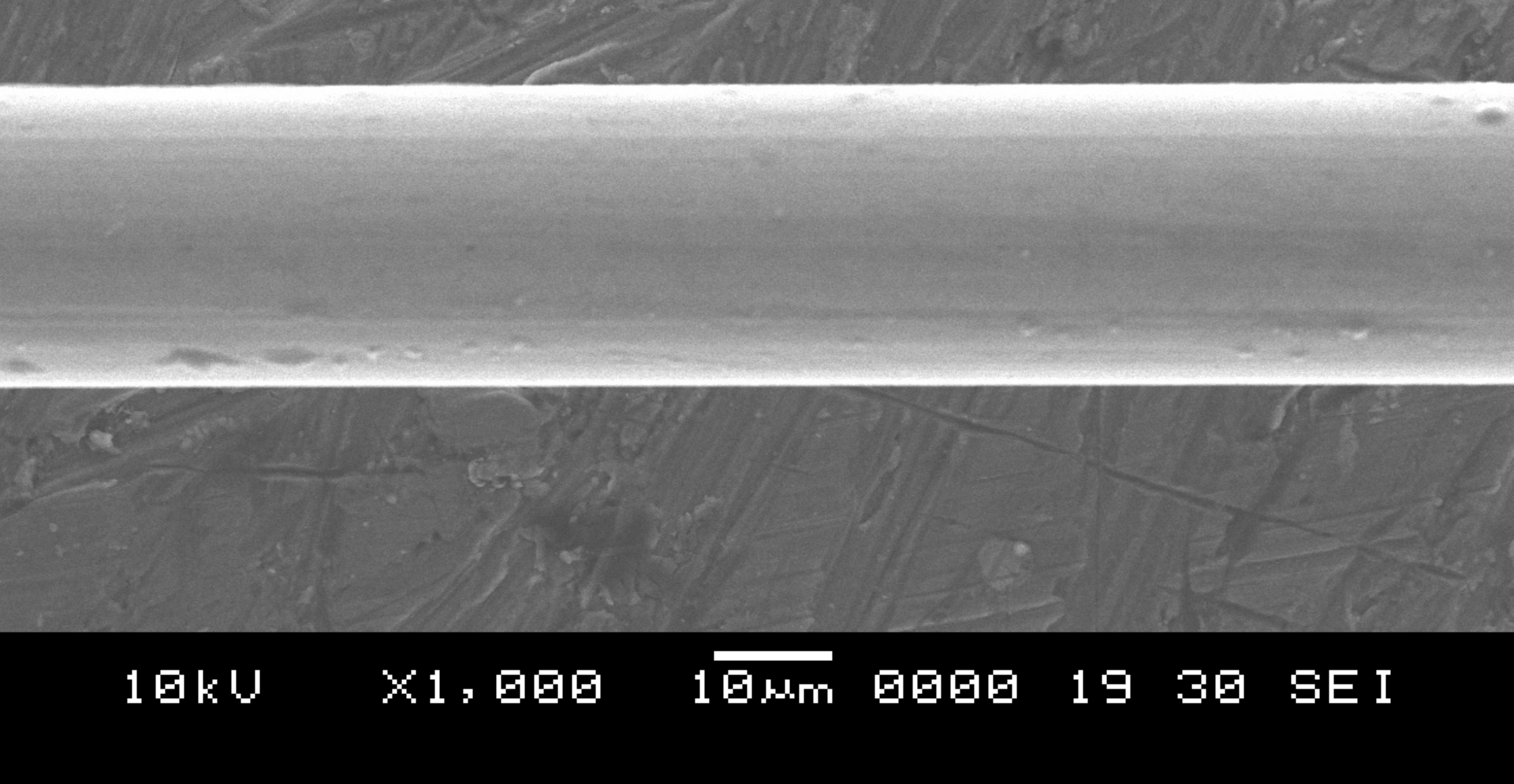}
\includegraphics[width=3.2in]{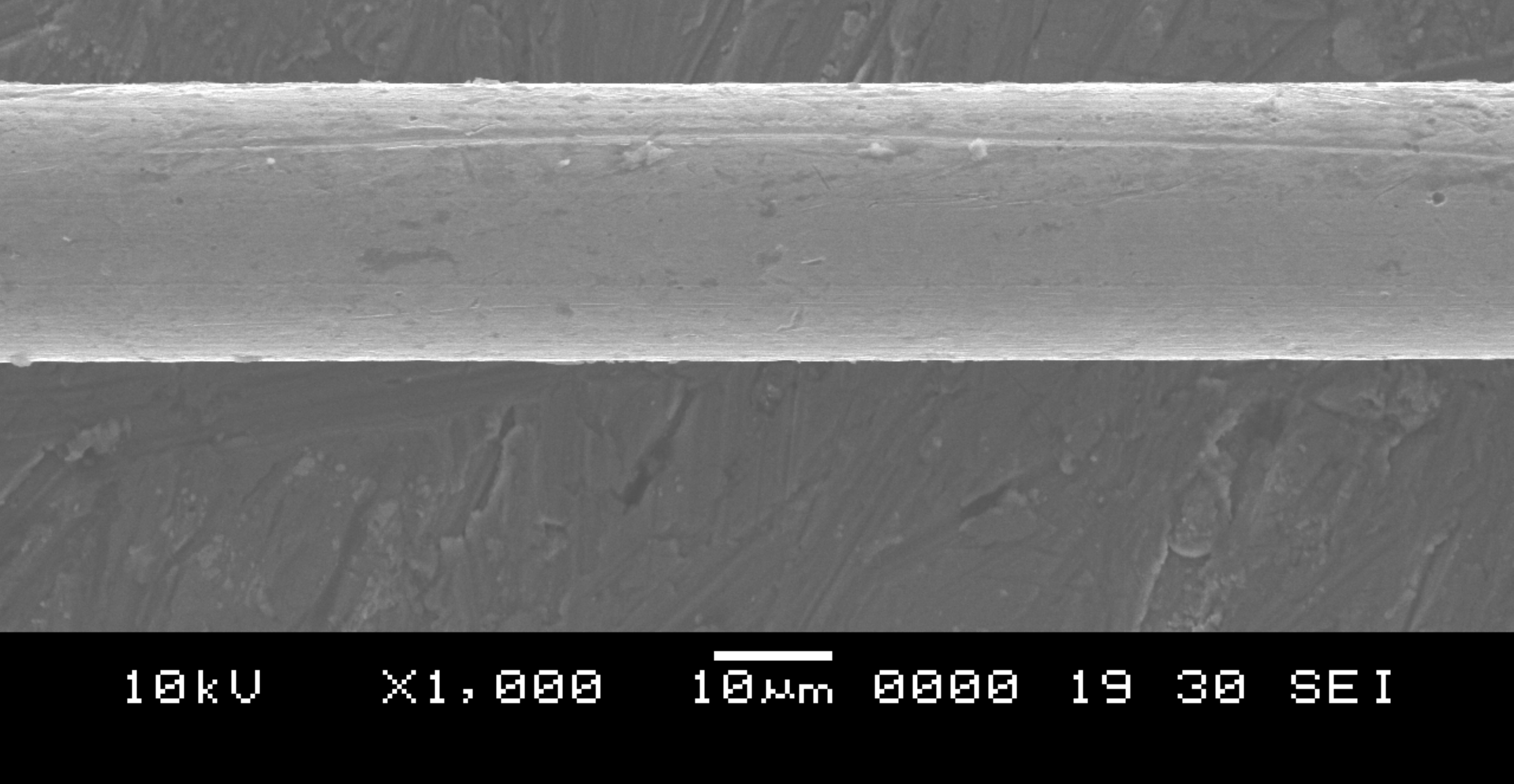}
\caption{Image of an originally 25-micron-diameter stainless steel wire before (left) and after (right) 60\,seconds of electropolishing in the same setup shown in Fig.~\ref{fig:epsetup}, which reduced the diameter about 2\,$\mu$m.}
\label{fig:SEMpicture}
\end{figure}

To test the uniformity resulting from electropolishing thin layers of stainless steel, a strand of 25-micron fine wire (California Fine Wire Co.) approximately two inches in length was electropolished in the same setup shown in Fig.~\ref{fig:epsetup} for sixty seconds, cleaned with ultra-pure water and isopropyl alcohol, and dried with compressed CO$_{2}$ gas. The wire was then measured at multiple points along its length using a scanning electron microscope and the GIMP image manipulation program (see Fig.~\ref{fig:SEMpicture}).  An unelectropolished wire was examined and analyzed for the control.  The unelectropolished wire showed a standard deviation $\sigma = 440$\,nm around its mean diameter $\mu = 25.08$\,$\mu$m.   The electropolished wire had $\sigma = 600$\,nm and $\mu = 22.74$\,$\mu$m.  The relatively large standard deviation on the unelectropolished wire may indicate that handling causing the wire to deform into an oval cross-section may be the dominant source of measured non-uniformity.  In any case, even for this relatively large thickness removed, the resulting uniformity is quite good.  In particular, the standard deviation $\sigma = 600$\,nm would be sufficient to provide $<10$\% gain variation in a drift chamber, sufficient for the proposed BetaCage~\cite{shutt_lrt2004,schneeLRT2006,ahmedLRT2010,protoJINST,LRT2013bunker}.

%%%%%%%%%%%%%%%%%%%%%%%%%%%%%%%%%%%%%%%%%%%%%%%%
%% BACKMATTER
%%%%%%%%%%%%%%%%%%%%%%%%%%%%%%%%%%%%%%%%%%%%%%%%

\begin{theacknowledgments}
This work was supported in part by the 
National Science Foundation (Grants No.\ PHY-0855525, PHY-0919278) and the 
U.S.\ Department of Energy (through Award Number DE-SCOO05054).
\end{theacknowledgments}



\bibliographystyle{aipproc}   % if natbib is available

\bibliography{schnee}

 \IfFileExists{\jobname.bbl}{}
 {\typeout{}
  \typeout{******************************************}
  \typeout{** Please run "bibtex \jobname" to obtain}
  \typeout{** the bibliography and then re-run LaTeX}
  \typeout{** twice to fix the references!}
  \typeout{******************************************}
  \typeout{}
 }

\end{document}